\documentclass[onecolumn]{revtex4}
\usepackage[T1]{fontenc}
\usepackage[latin1]{inputenc}
\usepackage{graphicx,color}
\usepackage{amsmath}
\usepackage{amssymb}
\usepackage{esint}

\makeatletter
\@ifundefined{textcolor}{}
{%
 \definecolor{BLACK}{gray}{0}
 \definecolor{WHITE}{gray}{1}
 \definecolor{RED}{rgb}{1,0,0}
 \definecolor{GREEN}{rgb}{0,1,0}
 \definecolor{BLUE}{rgb}{0,0,1}
 \definecolor{CYAN}{cmyk}{1,0,0,0}
 \definecolor{MAGENTA}{cmyk}{0,1,0,0}
 \definecolor{YELLOW}{cmyk}{0,0,1,0}
}

\@ifundefined{definecolor}
 {\usepackage{color}}{}
\usepackage{indentfirst}\usepackage{amsfonts}

\newcommand{\ei}{\mathrm{e}}
\newcommand{\tr}{\operatorname{tr}}
\newcommand{\ket}[1]{\left| #1 \right\rangle}
\newcommand{\bra}[1]{\left\langle #1 \right|}

\definecolor{gray}{rgb}{.4,.4,.4}
\definecolor{deepgreen}{rgb}{.1,.6,.3}

\makeatother

\begin{document}

\title{Parametric Competition in non-autonomous Hamiltonian Systems}

\author{L.~A.~M.~Souza$^{1}$}
\email{leonardoamsouza@ufv.br}

\author{J. G. P. Faria$^{2}$}

\author{M.~C.~Nemes$^{3}$}

\affiliation{$^{1}$Campus UFV-Florestal, Universidade Federal de Viçosa, CEP
35.690-000, phone/fax: +55-31-35363300, Florestal, Minas Gerais, Brazil}

\affiliation{$^{2}$Departamento de Física e Matemática, Centro Federal de Educação
Tecnológica de Minas Gerais, CEP 30510-000, Belo Horizonte, Minas Gerais, Brazil}

\affiliation{$^{3}$Departamento de Física, Instituto de Ciências Exatas, Universidade
Federal de Minas Gerais, CP 702, CEP 30123-970, Belo Horizonte, Minas
Gerais, Brazil.}

\begin{abstract}
In this work we use the formalism of chord functions
(\emph{i.e.} characteristic functions) to analytically solve
quadratic non-autonomous Hamiltonians coupled to a reservoir
composed by an infinity set of oscillators, with Gaussian
initial state. We analytically obtain a solution for the
characteristic function under dissipation, and therefore
for the determinant of the covariance matrix and the von Neumann entropy,  where the latter is the physical quantity of interest. We study in details two examples that are known to show dynamical squeezing and instability effects: the inverted harmonic oscillator and an oscillator
with time dependent frequency. We show that it will appear in both cases a clear competition between instability and dissipation. If the dissipation is small when compared to the instability, the squeezing generation is dominant and one can see an increasing in the von Neumann entropy. When the dissipation is large enough, the dynamical squeezing generation in one of the quadratures is retained, thence the growth in the von Neumann entropy is contained.
\end{abstract}
\maketitle

\section{Introduction}

The dynamics of quantum open systems has raised increasing
interest of physicists specially in the last decades \cite{zurek0,maximilian}:
it can be directly connected to the non-observance of quantum phenomena
in the classical world. The same phenomena which led Schroedinger to
discredit his own theory is directly connected to the linear structure
of the Hilbert space. Most of them have nowadays been observed and
the usual approach as to why they are not present in our everyday
life is to consider that quantum mechanics was first conceived for
closed systems and effects of the surrounding environment, when included,
tend to wash out quantum properties \cite{zurek0}.

This problem is however far from being a closed issue and the classical
limit of quantum mechanics is still a matter of enthusiastic debates
\cite{zurek0,maximilian,yaffe,haroche}.
In particular, the questions on quantum-to-classical transition acquire
a singular aspect in the case of quantum systems with
nonlinear or chaotic classical counterparts. If dissipation
is absent, it is expected that instabilities yield the fast
spreading of the wave function throughout
the phase space for such systems,
especially the macroscopic ones. Thus, an initially
well localized wave packet will soon
be fragmented throughout available regions of
the phase space, and coherent superpositions will
appear between the fragments, leading to
a rapid breakdown of the correspondence between classical
and quantum descriptions. Some authors \cite{zurek2,shiokawa,zurek3,habib,carlo}
advocate that the unavoidable interaction of a macroscopic system with
its environment is essential to prevent the appearing
of these quantum signatures yielded by inherent instabilities
exhibited by the unitary evolution. Notwithstanding, other authors \cite{wiebe} sustain that the coupling with an environment is not necessary because such quantum effects are so tiny that
they are not measurable, especially in the case of macroscopic
objects. This controversy only stresses the importance of the
study of the role played by instabilities in the question of
quantum-to-classical transition.


In the present contribution, we are concerned to the questions: what
happens if the unitary evolution, \emph{i.e.} the Hamiltonian of the
problem, may lead to instability? What role this instability
effects does play? Examples of application of non-autonomous
Hamiltonian systems can be found in a huge range of areas
of physics, in particular: in quantum optics, where a harmonic oscillator
with time dependent frequency is shown to generate squeezing \cite{agarwal_1,singh},
tunneling \cite{guo}, exact solutions for mathematical problems and toy models
\cite{karen}, parametric amplification \cite{agarwal_2},
quantum Brownian motion
\cite{dodonov}. Most of these works employs the model of the harmonic oscillator
with time dependent frequency. It is worth to mention that
this model is largely studied both in classical and quantum physics
and, as a merit, is amenable to analytical treatment.
In fact, the time independent Schroedinger equation for
the harmonic oscillator with time dependent frequency assumes the form of Hill
differential equation, which, in turn, is a particular form of Pinney
equation. Examples of Hill or Pinney equation in physics can be
found in studies on synchrotron accelerators \cite{courant},
anisotropic Bose-Einstein condensates \cite{BEC,BEC_cosmo}, Paul traps \cite{paul_traps},
and cosmological models of particle creation \cite{ray}.
Further, one of the first approaches to include dissipation in quantum
physics employed a class of time dependent Hamiltonians, known as
Caldirola-Kanai Hamiltonians \cite{dodonov,CK}.
Even in cosmology, in the inflationary era, when quantum effects are
supposedly important, studies using non-autonomous Hamiltonians, leading
to instabilities and squeezing effects are found \cite{cosmology}.
One then frequently uses non-autonomous unitary evolutions of the
same type, now modeling transitions between harmonic oscillators which give rise to particle
formation \cite{agarwal_1}. Quantum chaos and instabilities also
arise in recent experiments and theoretical models \cite{chaos},
rendering new perspectives to this interesting area in physics.
It is interesting to note that, due the features shared by both models,
some authors propose the Bose-Einstein condensates as a test bench
of some cosmological scenarios \cite{BEC_cosmo}.

Another interesting problem was 
raised by Zurek and collaborators \cite{zurek1} as to the rate of
entropy increase when the system of interest is coupled not to a reservoir
but to an unstable, two degrees of freedom system. In Ref.
\cite{karen} the authors analytically showed that, in fact, entropy
grows faster, but for that, chaos is not necessary (although sufficient).
Instability alone already reflects this physics. Also, more realistically,
as discussed in \cite{karen}, the potential modelling Paul-Penning
traps \cite{paul} has instability points which can be, to a certain degree, approximated
by an inverted oscillator. What happens to the well known physics described, if an environment
is added to the non-autonomous unitary dynamics? Can dissipation stop
the inevitable acceleration caused by instabilities?

A word about the formal mathematical approach to the problem is in
order: for autonomous systems, there are several possibilities to
solve a master equation. One of the frequently used and powerful tools
is that of Lie algebras of superoperators. Perhaps that is the reason
why there is not so much work devoted to the question of non-autonomous
systems evolving under nonunitary dynamics. As discussed above, however,
several interesting issues may be cleared, if one manages to formulate
the problem in appropriate language. In the present case, we will be considering single-mode Gaussian states. For these states, all we need are the second statistical moments or
the covariance matrix, which can be gotten very simply as derivatives
of the characteristic function (the Fourier transform of the Wigner function),
by taking the derivatives of this function at the origin \cite{ozorio_1,ozorio_2}.
Moreover, a very elegant theoretical method for Wigner functions and
nonunitary quadratic evolutions is given in Ref. \cite{ozorio_2}.
It involves several classical elements, rendering the physics of the
problem very transparent and the inclusion of nonunitary terms is
natural.

In section \ref{solution} we present an analytical solution for the
characteristic function, using the most general bilinear Lindbladian (for dissipative
reservoirs). We show our results for the inverted harmonic oscillator
(IHO) and for a non-autonomous harmonic oscillator (NAHO) with frequency
$\omega(t)=\omega_{0}\sqrt{1+\gamma t}$ in section \ref{results}
and in the last section we make our final remarks.

\section{Analytic solution for the Wigner and characteristic function}

\label{solution}

In this section we review some aspects concerning the evolution of single-mode Gaussian states under dissipation.
The literature is plenty of references on this subject (theory and applications)
\cite{PRA1, JPA_adesso, ozorio_1, ozorio_2, adessothesis, agarwal_1, schleich, piza, baseia1, baseia1993, baseia1997}.
To obtain our main result --- analytical solutions for non-autonomous Hamiltonians --- this section is,
although straightforward, useful.

\subsubsection{Unitary dynamics of single mode Gaussian states}

We can define a general form of the Hamiltonian part of the
equations of motion for both models studied in this work, namely,
the inverted harmonic oscillator (IHO) and the non-autonomous
harmonic oscillator. The Hamiltonian reads
\begin{equation}
	 \hat{H}(\hat{q},\hat{p},t)=\frac{\hat{p}}{2m}+\frac{1}{2}m\omega^{2}(t)\hat{q}^{2},\label{hamiltoneano2}
\end{equation}
where $ \hat{q}$ and $\hat{p} $ are position and linear momentum operators, respectively,
$m $ is the mass of the oscillator and $\omega(t)$ is a time-dependent frequency. If we
take $\omega_0 = \left|\omega\left(0\right)\right|$, the annihilation and creation operators for $t = 0$,
$\hat{a}$ and $\hat{a}^{\dagger}$, are given by
$\hat{a} = \sqrt{\frac{m\omega_0}{2\hbar}}\left(\hat{q}+i\frac{\hat{p}}{m\omega_0}\right)$ and
$\hat{a}^{\dagger} = \sqrt{\frac{m\omega_0}{2\hbar}}\left(\hat{q}-i\frac{\hat{p}}{m\omega_0}\right)$.
The Hamiltonian above can be written as \cite{baseia1993,baseia1997} \begin{equation} \hat{H}(\hat{a},\hat{a}^{\dagger},t)=\hbar\left[f_{1}(t)\left(\hat{a}^{\dagger}\hat{a}
	+\frac{1}{2}\right)+f_{2}(t)\left(\hat{a}^{\dagger2}+\hat{a}^{2}\right)\right],
\end{equation} where $
f_{1}(t)  =  \frac{\omega_{0}}{2}\left[\left(\frac{\omega(t)}{\omega_{0}}\right)^{2}+1\right]$ and $
f_{2}(t)  =  \frac{\omega_{0}}{4}\left[\left(\frac{\omega(t)}{\omega_{0}}\right)^{2}-1\right]$.

In order to establish the notation, we will first present single-mode
Gaussian states and its parameters, well known in the literature by
several methods. The initial state is
\begin{equation}
	\hat{\rho}(0)=\hat{D}(\alpha(0))\hat{S}(r(0),\phi(0))
	\hat{\rho}(\nu(0))\hat{S}^{\dagger}(r(0),\phi(0))\hat{D}^{\dagger}(\alpha(0)),
\end{equation}
where all the parameters are given by the first and second moments:
\begin{eqnarray}
	\alpha & = & \langle\hat{a}\rangle\nonumber \\
	\alpha^{*} & = & \langle\hat{a}^{\dagger}\rangle\nonumber \\
	\ei^{i\phi} & = & \sqrt{\frac{\sigma_{\hat{a}\hat{a}}}{\sigma_{\hat{a}^{\dagger}\hat{a}^{\dagger}}}}\nonumber \\
	\nu & = & \sqrt{\left(\sigma_{\hat{a}^{\dagger}\hat{a}}-\frac{1}{2}\right)^{2}
		-\sigma_{\hat{a}^{\dagger}\hat{a}^{\dagger}}\sigma_{\hat{a}\hat{a}}}-\frac{1}{2}\nonumber \\
	r & = & \frac{1}{4}\ln\left(\frac{\sigma_{\hat{a}^{\dagger}\hat{a}}-\frac{1}{2}
		+\sqrt{\sigma_{\hat{a}^{\dagger}\hat{a}^{\dagger}}
		\sigma_{\hat{a}\hat{a}}}}{\sigma_{\hat{a}^{\dagger}\hat{a}}
		-\frac{1}{2}-\sqrt{\sigma_{\hat{a}^{\dagger}\hat{a}^{\dagger}}\sigma_{\hat{a}\hat{a}}}}\right).
\end{eqnarray}
 In the equations above $\sigma_{\hat{a}^{\dagger}\hat{a}^{\dagger}}=
 \langle(\hat{a}^{\dagger})^{2}\rangle-\langle\hat{a}^{\dagger}\rangle^{2}$,
 $\sigma_{\hat{a}\hat{a}}=\langle(\hat{a})^{2}\rangle-\langle\hat{a}\rangle^{2}$,
 $\sigma_{\hat{a}^{\dagger}\hat{a}}=\langle\hat{a}^{\dagger}\hat{a}\rangle
 -\langle\hat{a}^{\dagger}\rangle\langle\hat{a}^{\dagger}\rangle+1$.
Those parameters are related to displacement ($\alpha$), squeezing
($r,\phi$) and ``impurity'' ($\nu$) of the state. In our study
the initial state will always be in this general single-mode Gaussian
form and, since the dynamics is quadratic, the state will evolve as
a single-mode \emph{Gaussian} state \cite{ozorio_2}.

One can study the state by analyzing the evolution of the parameters above, or the
covariance matrix (CM):
\begin{equation}
	\sigma=\left(
	\begin{array}{cc}
		\langle\hat{q}^{2}\rangle-\langle\hat{q}\rangle^{2} &
		\frac{1}{2}\langle\hat{q}\hat{p}+\hat{p}\hat{q}\rangle-\langle\hat{q}\rangle\langle\hat{p}\rangle\\
		\frac{1}{2}\langle\hat{q}\hat{p}+\hat{p}\hat{q}\rangle-\langle\hat{q}\rangle\langle\hat{p}\rangle &
		\langle\hat{p}^{2}\rangle-\langle\hat{p}\rangle^{2}
	\end{array}\right).
\end{equation}


\subsubsection{Wigner and Characteristic Functions --- Dissipationless case}

The Wigner function is defined as \cite{schleich}
\begin{equation}
	W(\vec{x})=\frac{1}{2\pi\hbar}\int dq'\left\langle q+\frac{q'}{2}\right|\hat{\rho}
	\left|q-\frac{q'}{2}\right\rangle \exp\left(-i\frac{pq'}{\hbar}\right),
	\label{wigner}
\end{equation}
where $\vec{x}=(p,q)$. It propagates ``classically'' for up to
quadratic dynamics \cite{ozorio_2}:
\begin{equation}
	\frac{\partial}{\partial t}W_{t}(\vec{x})=\{H(\vec{x}),W_{t}(\vec{x})\},
\end{equation}
where $\{f,g\}=\frac{\partial f}{\partial q}\frac{\partial g}{\partial p}-
\frac{\partial f}{\partial p}\frac{\partial g}{\partial q}$
is the classical Poisson bracket, and $H(\vec{x})=\vec{x}\cdot \hat{H} \vec{x}$.

One can write the propagated Wigner functions as \cite{arnold}:
\begin{equation}
W_{t}(\vec{x})=W_{0}(\mathbf{R}_{-t}\vec{x}),\label{evolucaowigner}
\end{equation}
where
\begin{equation}
\mathbf{R}_{t}=\exp(2 \boldsymbol{\Omega} \hat{H} t),\label{evolucaoR}
\end{equation}
 and $\boldsymbol{\Omega}$ is the symplectic form:
\begin{equation}
\boldsymbol{\Omega}=\left(
	\begin{array}{cc}
		0 & 1\\
		-1 & 0
	\end{array}\right).
\end{equation}

Note that \eqref{evolucaoR} reflects the evolution of a classical
Hamiltonian for points in phase space, \emph{i.e.}, given $H(\vec{x})$,
the time evolution of the variables is $\vec{x}_{t}=\mathbf{R}_{t}\vec{x}$,
or, explicitly
\begin{equation}
	\left(
	\begin{array}{c}
		p_{t}\\
		q_{t}
	\end{array}
	\right)=\left(
	\begin{array}{cc}
		\dot{v}_{t} & \dot{u}_{t}\\
		v_{t} & u_{t}
	\end{array}\right)\left(
	\begin{array}{c}
		p\\
		q
	\end{array}\right).\label{evolucaox}
\end{equation} This ``classical'' feature will become clear in the two models we studied in this work. The quantum behavior, for instance the dynamical squeezing generation due to the non-autonomous Hamiltonian, will be sustained by the non-unitary dynamics in some conditions, since the ``classicality'' will prevail through the ``quantumness'' for these cases.

For the Hamiltonian given by Eq. \eqref{hamiltoneano2},
both functions $u_{t}$ and $v_{t}$ obey the following equation \cite{agarwal_1}:
\begin{equation}
\ddot{\phi}_{t}+\omega^{2}(t)\phi_{t}=0.
\end{equation}
Here $\phi_{t}$ represents both $u_{t}$ or $v_{t}$. Two initial value
problems are defined with the above equation,  provided
the initial conditions $u\left(0\right)= 1$, $\dot{u}\left(0\right)= 0$ or
$v\left(0\right)= 0$, $\dot{v}\left(0\right)= 1$.
From equation \eqref{evolucaox}
one can see that
\begin{equation}
	\mathbf{R}_{t}=\left(
	\begin{array}{cc}
		\dot{v}_{t} & \dot{u}_{t}\\
		v_{t} & u_{t}
	\end{array}\right).\label{Rt}
\end{equation}

The characteristic function is the Fourier transform of the Wigner function \eqref{wigner} and it is given
by
\begin{equation}
	\chi(\vec{\xi})=\frac{1}{2\hbar\pi}\int d\vec{x}\exp\left(-\frac{i}{\hbar}
	\vec{\xi}\wedge\vec{x}\right)W(\vec{x}).
\end{equation}
The ``wedge'' product in the above equation is defined by
$\vec{\xi}\wedge\vec{x} = \xi_pq - \xi_qp$.
Since we will be working with single-mode Gaussian states, it is easy
to compute its initial ($t=0$) Wigner function as
\begin{equation}
W_{0}(V)=\frac{\exp\left(-\frac{1}{2}V\sigma^{-1}V^{T}\right)}{\pi\sqrt{\det\sigma}},
\end{equation}
where $V=(q,p)$ and $\sigma$ is the covariance matrix. Thus, the
characteristic function for the initial state is
\begin{equation}
\chi_{0}(\vec{\xi})=\frac{1}{2\pi}\int dV \exp\left[-i(\xi_{p}q-\xi_{q}p)\right]W_{0}(V).
\end{equation}
Evolving the Wigner function with Eq. \eqref{evolucaowigner}, one can
find the general solution of the characteristic function for the \emph{dissipationless
case}. Note the classical ingredient introduced by $\mathbf{R}_{t}$ and Eq. \eqref{evolucaox} in the solution.

\subsubsection{Wigner and Characteristic Functions --- Dissipative Case}

In this section we introduce nonunitary terms to the non-autonomous
dynamics considered above. In order to do this, we suppose the system
of interest coupled to a thermal bath at temperature $T$. The nonunitary contribution is given by
\begin{equation}
	\dot{\hat{\rho}}=\mathcal{L}\hat{\rho},
\end{equation}
with
\begin{eqnarray}
	\mathcal{L}\cdot & = & -i[\hat{H}\left(t\right),\cdot]
	+k(\bar{n}_{B}+1)(2\hat{a}\cdot\hat{a}^{\dagger}-\hat{a}^{\dagger}\hat{a}\cdot
	-\cdot\hat{a}^{\dagger}\hat{a})\nonumber \\
 	&  & +k~\bar{n}_{B}(2\hat{a}^{\dagger}\cdot\hat{a}-\hat{a}\hat{a}^{\dagger}\cdot
 	-\cdot\hat{a}\hat{a}^{\dagger}).\label{liouvilliano2}
\end{eqnarray}
Here, $\hat{H}\left(t\right)$ is the non-autonomous Hamiltonian presented in Eq.
\eqref{hamiltoneano2}, $k $ is a dissipation constant and $\bar{n}_{B}$ is the average number
of thermal excitations of the bath.

It is relatively simple, using the approach presented in this section, plus Gaussian
states and quadratic Hamiltonians, to obtain the characteristic function
\cite{ozorio_2}:
\begin{equation}
	\chi_{t}(\vec{\xi})=\chi_{0}(\vec{\xi}_{-t})
	\exp\left(-\frac{1}{2}\vec{\xi}\cdot\mathbf{M}(t)\vec{\xi}\right),
\end{equation}
where
\begin{equation}
	\mathbf{M}(t)=\sum_{j}\int_{0}^{t}dt'\ei^{2k(t'-t)}\mathbf{R}_{t'-t}^{T}
	(\mathbf{l}_{j}^{'}\mathbf{l}_{j}^{'T}+\mathbf{l}_{j}^{''}\mathbf{l}_{j}^{''T})
	\mathbf{R}_{t'-t},\label{mt}
\end{equation}
with
$j=1,2$. The variables $\vec{\xi}$ evolve in
time as
\begin{equation}
	\vec{\xi}_{t}=\ei^{kt}\mathbf{R}_{t}\vec{\xi}.\label{xit}
\end{equation}
For the specific case of equation \eqref{liouvilliano2} we have found
\begin{eqnarray*}
	\mathbf{l}_{1}^{'}=\left(
	\begin{array}{c}
		0\\
		\sqrt{k(\bar{n}_{B}+1)}
	\end{array}\right),~~
	\mathbf{l}_{1}^{''}=\left(
	\begin{array}{c}
		\sqrt{k(\bar{n}_{B}+1)}\\
		0
	\end{array}\right),
\end{eqnarray*}

\begin{eqnarray*}
\mathbf{l}_{2}^{'}=\left(\begin{array}{c}
0\\
\sqrt{k\bar{n}_{B}}
\end{array}\right),~~\mathbf{l}_{2}^{''}=\left(\begin{array}{c}
-\sqrt{k\bar{n}_{B}}\\
0
\end{array}\right),
\end{eqnarray*}
so that, in matrix \eqref{mt}: $\mathbf{l}_{1}^{'}\mathbf{l}_{1}^{'T}+\mathbf{l}_{1}^{''}\mathbf{l}_{1}^{''T}=k(\bar{n}_{B}+1)\mathbf{1}$
e $\mathbf{l}_{2}^{'}\mathbf{l}_{2}^{'T}+\mathbf{l}_{2}^{''}\mathbf{l}_{2}^{''T}=k\bar{n}_{B}\mathbf{1}$.

Notice that the apparent \emph{form} of the characteristic function
is very similar to the free case. The influence of the nonunitary
dynamics is contained in the matrix $\mathbf{M}(t)$ \eqref{mt} and
in the evolution of the variable $\vec{\xi}$ \eqref{xit}.
The term $\mathbf{R}_{t}$ contain the classical evolution. In order to obtain dissipative effects,
we have computed analytically $\mathbf{M}(t)$ for both studied cases.

The physics about the system is contained in the elements of the covariance
matrix (CM), since the CM completely define any Gaussian state. Dissipative effects are related to the determinant of the covariance matrix, $D(t)=\det\sigma$, which can be analytically calculated
from the derivatives of the characteristic function. The von Neumann entropy
for single-mode Gaussian states is completely defined by \cite{JPA_adesso, adessothesis}:
\begin{equation}
	S(t)=\left(\sqrt{D(t)}+\frac{1}{2}\right)\ln\left(\sqrt{D(t)}
	+\frac{1}{2}\right)-\left(\sqrt{D(t)}
	-\frac{1}{2}\right)\ln\left(\sqrt{D(t)}-\frac{1}{2}\right).
\end{equation} In the following section, we will obtain analytical results for the von Neumann entropy of two important examples, showing that is a clear competition between dissipative and instability effects.

\section{Results}

\label{results}

We are working with quadratic Hamiltonians \eqref{hamiltoneano2}, and as discussed in the
introduction, the time dependence in the oscillator frequency can generate squeezing \cite{agarwal_1,singh}.
Dissipative effects will be fully reflected by the von Neumann entropy or indirectly by $D(t)$: if
the state is isolated, its entropy will be constant; if the state is coupled to a thermal reservoir,
the von Neumann entropy will change in time \cite{leo_single}. In this section we show that a
competition between the dissipation constant and the non-autonomous frequency appear, and the
dissipation can sustain a possible increasing in the von Neumann entropy due to instability effects
(as dynamical squeezing generation). The general Hamiltonian (unitary part of the dynamics) in
this section is in the form given by \eqref{hamiltoneano2}: $\hat{H}(\hat{q},\hat{p},t)=\frac{\hat{p}}{2m}+\frac{1}{2}m\omega^{2}(t)\hat{q}^{2}.
$

\subsubsection{The Inverted Harmonic Oscillator (IHO)}

Let us consider first the toy model of the inverted harmonic
oscillator \cite{zurek1, iho} where $\omega(t)=i\omega_{0}$,
with $\omega_{0}$ a real constant. For the IHO, the matrix $\mathbf{R}_{t}$ is (where we have set $\omega_{0}=1$):
\begin{equation}
	\mathbf{R}_{t}=\left(
	\begin{array}{cc}
		\cosh t & \sinh t\\
		\sinh t & \cosh t
	\end{array}\right).\label{RtOHI}
\end{equation}
The matrix $\mathbf{M}(t)$ is:
\begin{equation}
	\mathbf{M}(t)=2k\left(\bar{n}_{B}+\frac{1}{2}\right)
	\int_{-t}^{0}dx\ei^{2kx}\left[\cosh2x\left(
	\begin{array}{cc}
		1 & 0\\
		0 & 1
	\end{array}\right)+\sinh2x\left(
	\begin{array}{cc}
		0 & 1\\
		1 & 0
	\end{array}\right)\right].
\end{equation}

\begin{figure}[!h]
\begin{center}
\includegraphics[scale=0.65]{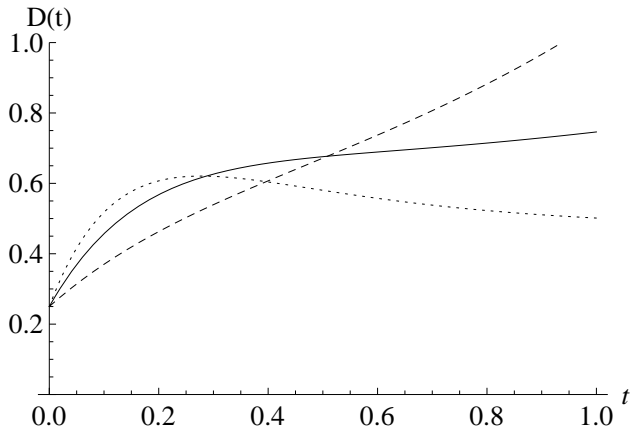} \caption{Determinant of the Covariance Matrix $D(t)$ as a function of time
for the IHO. Parameters: $\omega_{0}=1$, $r_{0}=1$, $\nu_{0}=0$,
$\bar{n}_{B}=0$. The dissipation constants are: $k=0.5$ (dashed), $k=1$
(solid) e $k=1.5$ (dotted).}\label{plotdet}\end{center}
\end{figure}

\begin{figure}[!htb]
\begin{centering}
\includegraphics[scale=0.65]{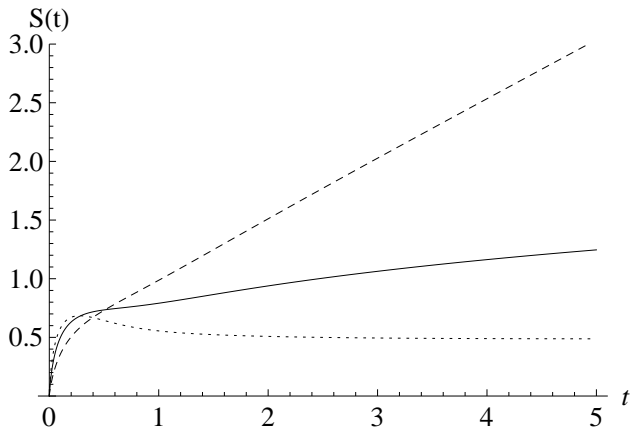} \caption{Von Neumann entropy as a function of time for the IHO. Parameters:
$\omega_{0}=1$, $r_{0}=1$, $\nu_{0}=0$, $\bar{n}_{B}=0$. The dissipation
constants are: $k=0.5$ (dashed), $k=1$ (solid) e $k=1.5$ (dotted).}\label{plotneuman}\end{centering}
\end{figure}

In Figure \eqref{plotdet} we show the determinant $D(t)$ for three
values of the dissipative constant $k$. Note that when $k$ is large
enough, compared with the frequency $\omega_{0}$, the determinant
goes to a constant value, larger than the initial one. The same behavior is
reflected in Figure \eqref{plotneuman}, where we show
the von Neumann entropy for this case.

In Ref. \cite{karen}, the authors showed that the coupling between a Gaussian
state in a harmonic oscillator ($\ket{\psi_{A}}$) with another Gaussian
state in an IHO ($\ket{\psi_{B}}$) will produce, in the the reduced
state $\rho_{A}=\tr_{B}\ket{\psi_{A}\psi_{B}}\bra{\psi_{A}\psi_{B}}$,
a linear increasing in the time evolution of von Neumann entropy. The authors argued
that the instability (and therefore the squeezing generation) of the
IHO will generate this effect. Here we can see that, if $k/\omega_{0}>1$
the dissipation will sustain the entropy growing, to a limit where
$S(t)$ becomes constant. Otherwise, if the dissipation is lesser
than the ``intensity'' of the instability (measured by $\omega_{0}$),
the entropy will increase monotonically, clearly tending to a linear increasing (as in \cite{karen}).
We argue here that the dissipation will suppress the squeezing generated by the IHO.

One can make a simple classical analogy in this case: suppose a simple rod which can spin up and down, by one of its end. If the rod is close to the bottom, it will ideally oscillates as an harmonic oscillator; but if the rod is in the top, and is released in a not so viscous fluid (as the air), it will oscillates and tend to a configuration totally different to the initial one (the rod in the top). But if the rod is released from the top in a viscous fluid (in the water or in a more exceeded example in a tar pit), the viscosity will retain the rod rotation. This classical behavior in the quantum IHO is pictured in the formalism given in section \ref{solution}, specially in the quantities $\mathbf{R}_{t}$ and $\mathbf{M}_{t}$.

\subsubsection{Simple model for dynamical squeezing generation}

\begin{figure}[!h]
\begin{centering}
	\includegraphics[scale=0.5]{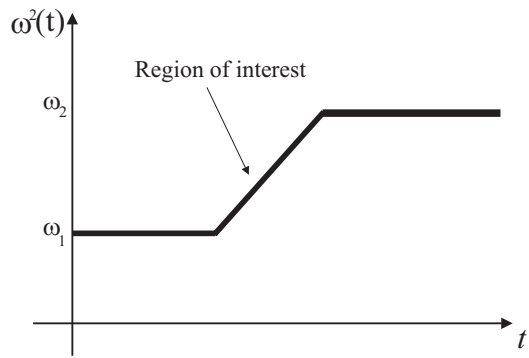}
	\caption{A pictorial image of the non-autonomous harmonic oscillator model.
	Our interest is to study the important
	region between the two harmonic oscillators, due to squeezing generation and instability effects.}\label{figuraosciladores}\end{centering}
\end{figure}

We consider now the following dynamics: two harmonic oscillators Hamiltonians
(autonomous evolution) with different natural frequency $\omega_{i}$,
separated each other by a non-autonomous Hamiltonian with modulated frequency
$\omega(t)$. We will be concerned with the part of the time dependence
which leads from the first to the second oscillators regime (Figure \eqref{figuraosciladores}).

In this model, when the state evolves trough the first to the second harmonic
oscillator, one can dynamically produce single-mode squeezing, and
therefore $\langle\hat{a}^{\dagger}\hat{a}\rangle\neq0$. We have used the following time dependence for the modulated frequency (as proposed in \cite{agarwal_1}):
\begin{equation}
	\omega(t)=\omega_{0}\sqrt{1+\gamma t}.
\end{equation}

For this time dependence, one can calculate the matrix $\mathbf{R}_{t}$
\eqref{Rt}, where the functions $u$ and $v$ satisfy ($u,v\rightarrow\phi_{\tau}$):
\begin{equation}
	\ddot{\phi}_{\tau}+\omega_{0}^{2}(1+\gamma\tau)\phi_{\tau}=0,
\end{equation}
for the following initial conditions: $u(0)=1$, $\dot{u}(0)=0$,
$v(0)=0$ and $\dot{v}(0)=1$. The solution is given by (where we
have set $\omega_{0}=1$):
\begin{equation}
	\phi_{\tau}=\text{Ai}\left[-\frac{1+\gamma\tau}{(-\gamma)^{2/3}}\right]C_{1}
	+\text{Bi}\left[-\frac{1+\gamma\tau}{(-\gamma)^{2/3}}\right]C_{2},
\end{equation}
where $C_{1}$ and $C_{2}$ are constants, $\text{Ai}$ and $\text{Bi}$
are the Airy functions and
\begin{equation}
	\tau=\omega_{0}t.
\end{equation}

Since we have the functions $u$ and $v$, we can write the matrix
$\mathbf{R}_{t}$ and compute $\mathbf{M}(t)$:
\begin{equation}
	\mathbf{M}(t)=2k\left(\bar{n}_{B}+\frac{1}{2}\right)
	\int_{-t}^{0}dx\ei^{2kx}\mathbf{R}_{x}^{T}\mathbf{R}_{x}.
\end{equation}
Now we are able to study the physical quantity of interest, the von
Neumann entropy.

\begin{figure}[!h]
\begin{centering}
	\includegraphics[scale=0.65]{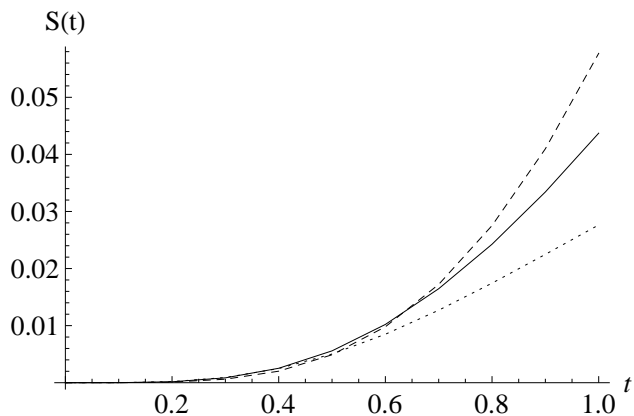}
	\caption{Time evolution of the von Neumann entropy for a initially pure Gaussian
	state ($\nu_{0}=0$) evolving in an unitary Hamiltonian with frequency
	$w(t)=w_{0}\sqrt{1+\gamma t}$ coupled to a thermal reservoir with
	$\bar{n}_{B}=0$. Parameters: $\omega_{0}=1$, $r_{0}=0$, $\gamma=1$.
	The dissipation constants are: $k=0.5$ (dashed), $k=1.0$ (solid)
	and $k=1.5$ (dotted).}\label{plotneuman1}\end{centering}
\end{figure}

\begin{figure}[!h]
\begin{centering}
	\includegraphics[scale=0.65]{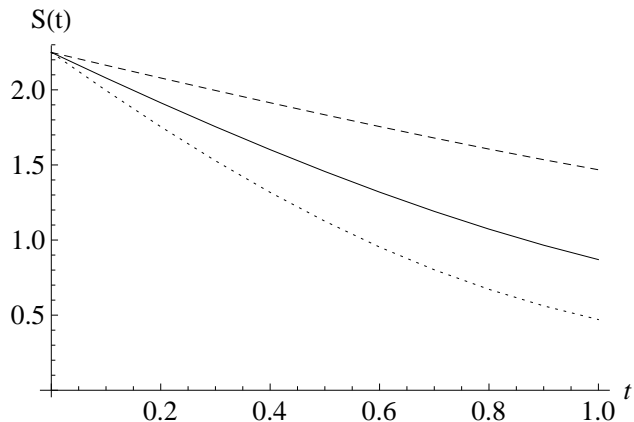} 	
	\caption{Time evolution of the von Neumann entropy for a initially non-pure
	Gaussian state ($\nu_{0}=3$) evolving in an unitary Hamiltonian with
	frequency $w(t)=w_{0}\sqrt{1+\gamma t}$ coupled to a thermal reservoir
	with $\bar{n}_{B}=0$. Parameters: $\omega_{0}=1$, $r_{0}=0$, $\gamma=1$.
	The dissipation constants are: $k=0.5$ (dashed), $k=1.0$ (solid)
	and $k=1.5$ (dotted).}\label{plotneuman2}\end{centering}
\end{figure}

For this model, we obtain the results given in figures \eqref{plotneuman1}
and \eqref{plotneuman2}, respectively the von Neumann entropy for
a Gaussian state evolving thought an unitary dynamics with frequency
$\omega(t)=\omega_{0}\sqrt{1+\gamma t}$ for a initially pure state
($\nu_{0}=0$) and for a thermal state ($\nu_{0}=3$), both coupled
to a reservoir at zero temperature. We have used those values of $\nu$ based on \cite{leo_single},
since in this work the authors showed that there is a maximum value of $\nu$ for which the state
present visible squeezing. Note the competition between unitary
and nonunitary effects: the latter will reduce the former always.
The amount of the squeezing suppression is governed by
$\frac{k}{\omega_{0}}$.

This example can be used for various purposes, from Paul traps \cite{paul_traps} to
cosmological models concerning the initial Universe \cite{cosmology}. Since the dynamical
squeezing generation is suppressed by the dissipation, one can conjecture, for example, that the
increasing in the average number of particles (also generated dynamically:
$\langle\hat{a}^{\dagger}\hat{a}\rangle \neq 0$) will be contained by the dissipation.
This application is direct implications for inflation models in cosmology related to particle and anti-particle generation.

\section{Conclusion}

In the present contribution, we have shown that the Wigner formalism
as constructed in Ref. \cite{ozorio_2} is most adequate for handling non-autonomous
dissipative systems. This calculation facility comes mainly from working
with the characteristic function and Gaussian states.

We have been able to show that squeezing generation as observed by
time dependent frequencies of the harmonic oscillator may be limited
by the process of decoherence. As to instabilities here simulated
simple by two examples ($\omega=i\omega_{0}$ and $\omega(t)=\omega_{0}\sqrt{1+\gamma t}$)
a clear competition between instability and dissipation appears. Interestingly
enough, for large dissipation (in comparison with the natural frequency
$\omega_{0}$), the time evolution of the quadratic variances, or
the entropy, reaches an asymptotic limit.

\textbf{Acknowledgments}

LAMS thanks the Brazilian agencies Fundação Arthur Bernardes
(FUNARBE -- FUNARPEX III-2013), Fundação de Amparo à Pesquisa de Minas Gerais (FAPEMIG)
and Conselho Nacional de Desenvolvimento Científico e Tecnológico
(CNPq -- 470131/2013-6) for financial support.
JGPF acknowledges the support from CNPq (grants
486920/2012-7 and 306871/2012-2) and from CEFET/MG (PROPESQ
program -- grant 10122\_2012). MCN and JGPF were partially
supported by Institulo Nacional de Ciência e Tecnologia em
Computação Quântica (INCT-IQ). JGPF and LAMS would like to
dedicate this work in memory  of M. C. Nemes.

\end{document}